# Spin(ing) into the classroom:
## Quantum spin activities for Year 6—10 physics

Kyla Adams[a], Anastasia Lonshakova[a], David Blair[a], David Treagust[b], Tejinder Kaur[a]

[a] School of Physics, The University of Western Australia, Perth, Australia.
[b] School of Education, Curtin University, Perth, Australia.

## Abstract

Quantum science is in the news daily and engages students' interest and curiosity. A fundamental quantum science concept that underpins medical imaging, quantum computing and many future technologies is quantum spin. Quantum spin can explain many physical phenomena that are in the lower secondary school curriculum, such as magnetism and light, making its inclusion a great motivator for students. Here we present an activity sequence for teaching quantum spin in the classroom using spinning tops and gyroscopes to highlight the common properties of classical angular momentum and quantum spin. These toys can provide an easily understood window to the quantum world for lower secondary school students. Students who have engaged in these activities reported enjoying the content and appreciating its relevance.

## What is quantum spin?

The newest frontiers in technology are based on quantum science. This is evident from the 2023 Nobel Prizes in physics and chemistry which are both based on quantum technologies (Laurence, 2023; Niranjan, 2023). The demand for a quantum-informed workforce will grow over the coming years, with start-ups, government organisations and universities all developing quantum technologies (Kaur & Venegas-Gomez, 2022). Future jobs will require a range of quantum science understanding, from a broad appreciation of the concepts to expert knowledge. The benefits of teaching quantum science extend to the wider community by increasing science literacy, an appreciation for the nature of science and improved scientific reasoning skills (Park et al., 2019).

Quantum science covers a wide range of concepts, not only those behind future technologies like quantum computing, but also school curriculum concepts like light, magnetism, and energy transformations. One of the quantum concepts that is fundamental to our quantum understanding is *quantum spin*. Quantum spin is the term given to the quantised angular momentum that is intrinsic to particles. Angular momentum and quantum spin share similarities in their definitions and interactions with the world. Therefore, classical angular momentum can be a way to introduce students to quantum spin.

Motivated by the scientific relevance of quantum spin to our lives and perceived student interest in modern physics, we developed an activity-based teaching approach that provides students with an understanding of quantum spin. The activity sequence was developed with students in two interactive workshops as part of a flagship 'Quantum Girls' trial program.

Quantum Girls is a program designed to introduce quantum computing and science to students in Years 5-12.

The proposed teaching approach described in this paper reflects the methods developed by the Einstein-First research group. Einstein-First has spent the past 10 years working with teachers and students to develop ways to modernise physics teaching. This work has established a process for creating learning sequences that can be used by any teacher in Australian schools and around the world (Kersting & Blair, 2022). Over the past decade, there has been increasing interest in the inclusion of modern physics into the curricula (Henriksen et al., 2014; Stadermann et al., 2019). Quantum topics can motivate and encourage students, and spin is one fundamental concept of interest.

Teaching quantum spin is not new. Quantum spin is taught in university quantum mechanics classes using abstract mathematics only accessible to specialists, with some universities taking a 'spin-first' approach (Sadaghiani, 2016). What is new is using classical spin, with the appropriate scaffolding, to introduce quantum spin to lower secondary school students. Teaching quantum spin allows us to introduce a new range of concepts that can be taught to students, from magnets to light – topics that excite young minds (Angell et al., 2004).

## Background for teachers

Quantum spin is quantised angular momentum. It is linked to the classical angular momentum we see every day through observable outcomes but has important differences. For historical reasons quantised angular momentum was called 'spin', and the phrase has stuck.

Quantum spin was first observed in experiments in 1922 by Stern and Gerlach (Castelvecchi, 2022; Gerlach & Stern, 1922). Who fired silver atoms through a magnetic field. Originally atoms were thought to mimic a bar magnet, because of this, Stern and Gerlach predicted that the atoms would have random deflections as the atoms would be arranged in random directions. Passing these atoms through the magnetic field would then result in a single line of atoms in line with the magnetic field (see left-hand side of Figure 1). Instead, they saw two distinct groups as shown in the right-hand side of Figure 1. This experiment eventually led to the full description of quantum spin which is directly connected to magnetism.

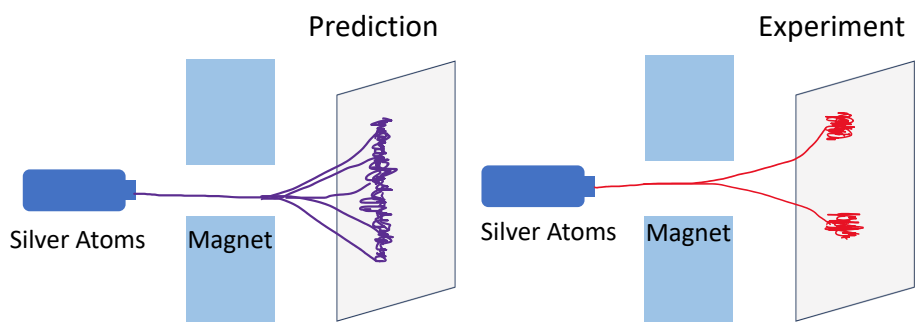

*Figure 1: The predicted (left) and experimental (right) results of the Stern-Gerlach experiment.*

To appreciate the similarities between quantum spin and classical angular momentum we need to take a step back to classical *angular momentum*.

Angular momentum is a rotating form of linear momentum. Linear momentum is a combination of the mass and speed of a moving object. Objects with more momentum are harder to stop. They have more energy. Angular momentum is the difficulty to stop rotational motion. Angular momentum can be broken down into two forms. One is orbital angular momentum, this is the angular momentum of an object moving about an origin, like the Earth orbiting the sun. The other is spin angular momentum, which is the angular momentum of an object about itself, like the Earth spinning on its axis.

Both classical and quantum angular momentum have these two forms. Classically we can see spin angular momentum in a spinning top. Quantum spin is similar, except that it is quantised, and is unstoppable. In quantum mechanics we observe spin through experiments like the Stern-Gerlach experiment and see the large-scale effects in magnetism. Orbital angular momentum also is quantised and allows us to describe electron shell types but is beyond the scope of this article. For simplicity, from now on we will only refer to the spin form of angular momentum.

Quantum spin can generate classical spin angular momentum because they exhibit the same properties. Table 1 compares classical and quantum spin. As can be seen, there are many similarities. These similarities enable classical spin to be used to introduce the concepts of quantum spin, provided it is introduced appropriately. The differences provide a fascinating window to the quantum world.

*Table 1: Comparison of properties of classical and quantum spins.*

| Property | Classical Spin | Quantum Spin |
|---|---|---|
| Can add | ☑ | ☑ |
| Can cancel | ☑ | ☑ |
| Can precess | ☑ | ☑ |
| A type of Angular momentum | ☑ | ☑ |
| Fixed, unique values | ☒ | ☑ |
| Used to classify fundamental particles | ☒ | ☑ |
| Fully described by classical arguments | ☑ | ☒ |
| Can be used to explain classical observations | ☑ | ☑ |

## Particle classifications

Since the Stern-Gerlach experiment, scientists around the world have quantified the spin of the particles that make up atoms. An electron has the same spin as a proton and high energy photons have the same spin as low-energy infrared photons. Quantum spin is defined by a unit called the modified Planck's constant ($\hbar = \frac{h}{2\pi}$). Electrons, protons, and neutrons all have the exact same spin value ($\frac{1}{2}\hbar$), while photons have exactly double that spin ($1\,\hbar$).

Every fundamental particle has so far been found to have either spin-$\frac{1}{2}\hbar$ (fermions), or spin-$1\hbar$ (bosons), except the Higgs particle that has spin zero. These classifications make it easier to create simple representations of quantum spin using arrows. Fermions are represented by an arrow that is half the length of an arrow for a boson. By convention, the direction of the spin arrow is represented by the right-hand rule (See Figure 2). Spin up is equal to positive spin values ($+\frac{1}{2}\hbar$), and spin down is equal to negative values ($-\frac{1}{2}\hbar$). Both classical and quantum spin can be represented by these arrows as they follow the same angular momentum rules. Classical spin arrows can have any length. As discussed later, both classical and quantum spins can flip between positive and negative.

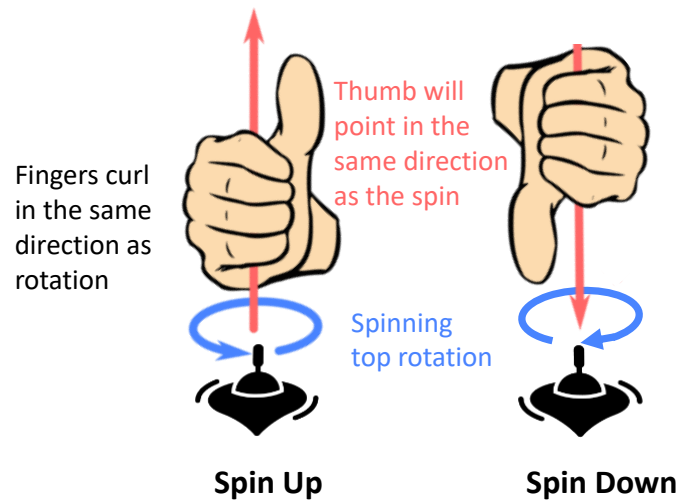

*Figure 2: Right-hand rule: when the fingers curl in the same direction as the rotation and the thumb is pointing up, we say the top has 'spin up' (Image adapted from Wikimedia Commons: SVGguru).*

## Classical spin from quantum spin

Determining the spin properties of fermions and bosons required new experiments, and new interpretations of old ones. Photon spin was first determined in 1933 (Beth, 1936). In an experiment a beam of polarised light was used to transfer angular momentum to a disk-shaped optical plate. The experiment measured the large-scale transfer of photon quantised spins to the plate resulting in a measurable classical spin. This outcome was possible as the polarised light meant all photons had the same spin direction, they could add together to a larger spin.

One of the first observations of electron spin was made by Albert Einstein and Wander de Haas in 1915. They conducted an experiment on the effect of a magnetic field on a hanging iron rod. With no magnetic field the rod remained stationary. This is a result of the random alignment of electron spins. When the magnetic field is turned on, the large number of small electron spins add up to create an overall change in angular momentum. The rod gains a classical spin. At that time, the researchers did not connect their observations to spin of particles, but instead to the connections of angular momentum and magnetism more generally. A video demonstration of this effect can be found here (UMDemoLab, 2018). This experiment is one of many examples where we see quantum angular momentum resulting in classical angular momentum.

## Quantum spin and the science curriculum

Based on Version 9.0 of the Australian curriculum, from as early as Year 5, students are expected to start using particle theory to describe and explain different phenomena (e.g. Year 5 Chemical science elaboration: *explore through guided discussion ideas about what is between particles*)(ACARA, 2023). By Year 10 students are investigating atoms in detail, from electron orbital structures to differences between metals and non-metals. Magnetism is another concept that can be found throughout the Australian curriculum, but not why magnetism exists. Quantum spin explains magnetism, and why we can magnetise some objects, but not others.

Quantum spin is not explicitly mentioned in the curriculum, but it underpins many learning statements and can be easily incorporated into the classroom under the ACARA science curriculum elaborations. Curious students are always asking 'why?'. By teaching spin, we can provide answers and keep students engaged and interested. Magnetism as noted earlier is one such example where quantum spin can answer the 'why' questions.

## Classroom activities

Quantum spin and classical spin have similar observable outcomes and definitions. Even with the available technology students are unable to directly see a singular atom or particle spin. Instead, by using activities we can show the similarities between classical spin and quantum spin. Here we present the activity sequence for teaching quantum spin in the classroom. Table 2 provides a summary.

### Spin direction

Defining spin direction with the right-hand rule (Figure 2) is important as it creates a common language for students. The first activity we present allows students to use spinning tops to practice defining spin directions (Figure 3).

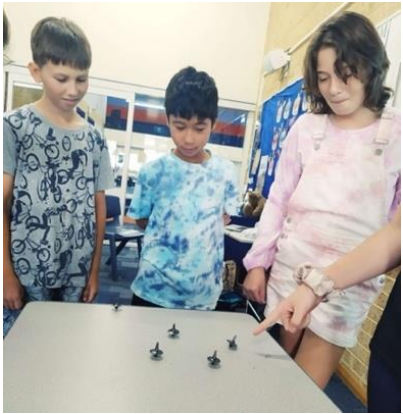

*Figure 3: Students defining spin direction using spinning tops.*

### Spin precession

Spin precession occurs when an external force is applied to a spin. In the case of a toy gyroscope, gravity acts to destabilise the spin, leading to motion in the perpendicular direction. This causes motion causes the spin arrow to precess as shown by the green arrow in Figure 4 (bottom).

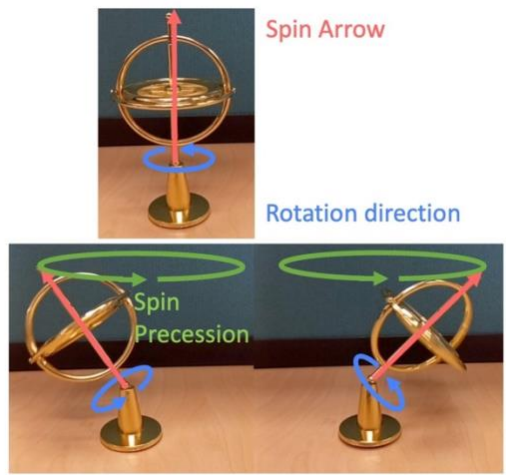

*Figure 4: Top: A gyroscope with no spin precession. The spin arrow (red) points directly upwards. Bottom: A gyroscope with spin precession (green) at two different points in time. The rotation direction remains the same.*

The spin arrow moves like a wobbling spinning top. Both classical and quantum spins can precess in this way. Students use gyroscopes and spinning tops to observe spin precession because of gravity.

Figure 4 shows a gyroscope with no precession (top) and two examples of a gyroscope with precession (bottom). We present precession as an observational fact, rather than the detail above as the mathematics is not accessible to this age group.

## Spin flip

A spin flip always means having a change in angular momentum direction. Some classically spinning objects give the illusion of a spin flip, like spinning a hard-boiled egg, but the spin arrow does not change (Figure 5). A common example of classical spin flip is flipping a gyroscope. When a spinning gyroscope is turned upside-down the spin flips. This requires you to apply forces to the gyroscope, hence supplying energy. When you do this there is also a transfer of angular momentum to the person doing the flipping. To flip the quantum spin of an electron you can hit the particle with photon energy.

The photon transfers both spin and energy. The electron spin can be flipped from $-\frac{1}{2}\hbar$ to $+\frac{1}{2}\hbar$ because the photon has a spin of 1. Spin flip is a result of both quantised energy and spin addition. Electron spin flips are important in radio astronomy as they allow astronomers to map hydrogen gas throughout the universe.

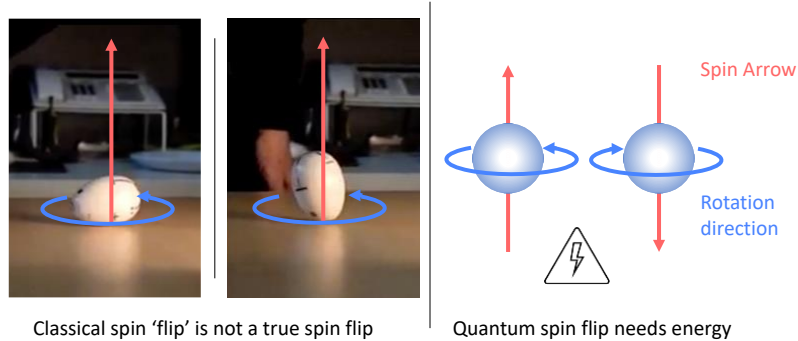

Classical spin 'flip' is not a true spin flip | Quantum spin flip needs energy

*Figure 5: A spinning hard-boiled egg will not change in spin direction, but it does change its orientation. A quantum spin flip has a change in spin direction, but only when energy is given (See Cross, 2017- for video).*

## Adding spin

Quantum spin arrows can add head to tail, and the shortest path from beginning to end is called the resultant. We call this the Math of Arrows (also known as vector addition, more information can be found in this video (Einstein-First, 2022)). Quantum spins add in the same way. Students can gain an intuition for adding spin using stacking spinning tops. Students playfully experiment with the spinning tops and find ways to add angular momentum (Figure 6).

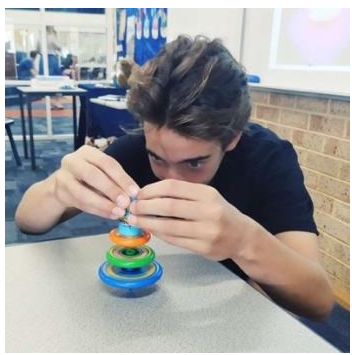

*Figure 6: A student stacking spinning tops.*

### Using quantum spin

A common medical imaging technique, Magnetic Resonance Imaging (MRI) uses spin precession, spin flips and spin addition to create a map of the inside of a human body. Hydrogen atoms make up ~ 63% of the total atoms in the human body (Shraga, 2020). Conveniently, hydrogen nuclei contain a single proton with spin-$\frac{1}{2}\hbar$ . When proton spins are exposed to a magnetic field, they will flip, like a compass aligning with north. These spins add, creating a net magnetism. This spin alignment is not exact. They precess at a frequency of approximately 60 MHz, which is of the order of short-wave radio or analogue tv transmissions (30-300MHz). When hit with the radio pulse, the spins will absorb the energy, and the spin alignment will change. Then, the spins will each lose energy in the form of a radio wave and realign with the magnetic field. We record the change in energy (an observed effect of angular momentum) to 'see' inside the body. Manipulating quantum spin has allowed us to develop many new technologies like MRI's. See this video for some useful animations (Johns Hopkins Medicine, 2022).

We have found that the activities described in Table 2 help students gain an appreciation of quantum spin, and applications like MRI's give them the opportunity to see the direct relevance of spin. We acknowledge that these activities on their own are not novel (e.g. see (Cross, 2017)). What is new here is the use of the activities combined with interpretations and analogies to uncover the relevant quantum concepts. This program has been taught to lower secondary school students, and serves as a gateway to the concepts of quantum physics (Popkova et al., 2023).

*Table 2: Recommended activity structure, relevant real-world applications, and curriculum links.*

| ACTIVITY | CONCEPT | REAL WORLD CONTEXT | CURRICULUM EXAMPLE |
|---|---|---|---|
| Spinning tops | Spin direction | Right-hand rule defines the spin | Year 3 Mathematics onwards |
| Gyroscopes | Spin precession | MRI's (see below), Earth axial precession | Year 6 Earth and Space sciences |
| Spinning hardboiled egg | Limits of classical spin flip | Quantum computing, Astronomy | Year 7 Physical sciences |
| Stacking tops | Adding spin | Magnetism, Atoms | Year 4 Physical sciences Year 5 Chemical sciences |

### Student reactions

The activities outlined here were initially tested on *World Quantum Day* with 20 participants (ages 10-15). It was later delivered to another 23 participants (ages 8-14) in the *Maths for Einstein's Universe* program which focused on the mathematical representation of physical processes that require a vectors description (e.g. forces, light and spin) (Popkova et al., 2023).

Regardless of age and program type, most students were interested in quantum spin and often mentioned the spinning top activity. After the programs, students generally saw quantum science as the future of technology, fundamental to our lives and a way to help us understand the world around us.

## Conclusion

Quantum technologies are our future. The basics behind these technologies should be taught to students the same way as we currently teach about energy and magnetism, which were once the concepts underpinning emerging technologies. Students are interested in quantum spin; they do not find the concept difficult when it is taught in a structured way as described in this paper with interactive activities. Students also recognise how learning about spin is important to our world. These activities can easily be incorporated into the classroom for students of any age and linked to many of the elaborations within the Australian curriculum.

## Acknowledgements

This work was supported by the Australian Research Council Linkage Grant LP 180100859. The research was carried out under the University of Western Australia Ethics approval number 2019/RA/4/20/5875. We thank the students for their participation in the workshops that allowed us to develop the work presented here. Photos have been reproduced with permission.

## REFERENCES

ACARA. (2023). *V9 Australian Curriculum*. Australian Curriculum, Assessment and Reporting Authority. https://v9.australiancurriculum.edu.au/

Angell, C., Guttersrud, Ø., Henriksen, E. K., & Isnes, A. (2004). Physics: Frightful, but fun. Pupils' and teachers' views of physics and physics teaching. *Science Education*, *88*(5), 683–706. https://doi.org/10.1002/sce.10141

Beth, R. A. (1936). Mechanical detection and measurement of the angular momentum of light. *Physical Review*, *50*(2), 115–125. https://doi.org/10.1103/PhysRev.50.115

Castelvecchi, D. (2022). The Stern–Gerlach experiment at 100. *Nature Reviews Physics*, *4*(3), 140–142. https://doi.org/10.1038/s42254-022-00436-4

Cross, R. (2017, December). *Egg2.MOV*. http://physics.usyd.edu.au/~cross/Egg2.MOV

Einstein-First (Director). (2022, August 1). *MF 1.1—Tug of War—The Math of Arrows*. https://youtu.be/NA4aPCVIdck?si=2PsVnjSL_wAW587Q

Gerlach, W., & Stern, O. (1922). Der experimentelle Nachweis der Richtungsquantelung im Magnetfeld. *Zeitschrift for Physik*, *9*(1), 349–352. https://doi.org/10.1007/BF01326983

Henriksen, E. K., Bungum, B., Angell, C., Tellefsen, C. W., Frågåt, T., & Bøe, M. V. (2014). Relativity, quantum physics and philosophy in the upper secondary curriculum: Challenges, opportunities and proposed approaches. *Physics Education*, *49*(6), 678–684. https://doi.org/10.1088/0031-9120/49/6/678

Johns Hopkins Medicine (Director). (2022, July 6). *MRI Physics | Magnetic Resonance and Spin Echo Sequences—Johns Hopkins Radiology*. https://www.youtube.com/watch?v=jLnuPKhKXVM

Kaur, M., & Venegas-Gomez, A. (2022). Defining the quantum workforce landscape: A review of global quantum education initiatives. *Optical Engineering*, *61*(08). https://doi.org/10.1117/1.OE.61.8.081806

Kersting, M., & Blair, D. (Eds.). (2022). *Teaching Einsteinian Physics in Schools: An essential guide for teachers in training and practice* (First published). Routledge, Taylor & Francis Group.

Laurence, M. (2023, October 5). *Nobel prize in chemistry awarded for 'quantum dot' technology that gave us today's high definition TVs*. The Conversation. https://theconversation.com/nobel-prize-in-chemistry-awarded-for-quantum-dot-technology-that-gave-us-todays-high-definition-tvs-214976

Niranjan, S. (2023, October 4). Making 'movies' at the attosecond scale helps researchers better understand electrons − and could one day lead to super-fast electronics. *The Conversation*. https://theconversation.com/making-movies-at-the-attosecond-scale-helps-researchers-better-understand-electrons-and-could-one-day-lead-to-super-fast-electronics-214931

Park, W., Yang, S., & Song, J. (2019). When Modern Physics Meets Nature of Science: The Representation of Nature of Science in General Relativity in New Korean Physics Textbooks. *Science & Education*, *28*(9–10), 1055–1083. https://doi.org/10.1007/s11191-019-00075-9

Popkova, A., Adams, K., Boublil, S., Choudhary, R. K., Horne, E., Ju, L., Kaur, T., McGoran, D., Wood, D., Zadnik, M., Blair, D. G., & Treagust, D. F. (2023). Einstein-First: Bringing children our best understanding of reality. *The Sixteenth Marcel Grossmann Meeting*, 2438–2452. https://doi.org/10.1142/9789811269776_0194

Sadaghiani, H. R. (2016). Spin First vs. Position First instructional approaches to teaching introductory quantum mechanics. *2016 Physics Education Research Conference Proceedings*, 292–295. https://doi.org/10.1119/perc.2016.pr.068

Shraga, A. (2020, April 1). *The Body's Elements*. https://davidson.weizmann.ac.il/en/online/orderoutofchaos/body%E2%80%99s-elements

Stadermann, H. K. E., Van Den Berg, E., & Goedhart, M. J. (2019). Analysis of secondary school quantum physics curricula of 15 different countries: Different perspectives on a challenging topic. *Physical Review Physics Education Research*, *15*(1), 010130. https://doi.org/10.1103/PhysRevPhysEducRes.15.010130

UMDemoLab (Director). (2018, May 7). *5H60.10—Einstein De Haas Effect*. https://www.youtube.com/watch?app=desktop&v=qFkW0PHhXcY